\documentclass[preprint,12pt]{elsarticle}

\usepackage{graphicx}
\usepackage{amssymb}
\usepackage{amsmath}
\usepackage{framed} 
\usepackage{bm}

\usepackage{xcolor}
\usepackage[margin=2.5cm]{geometry}
\usepackage[%
    colorlinks=true,%
    hyperindex,%
    plainpages=false,%
    bookmarksopen,%
    bookmarksnumbered]{hyperref}

\usepackage{tabularx}
\usepackage{booktabs} 

\makeatletter
\AtBeginDocument{\def\@citecolor{blue}}
\makeatletter
\def\ps@pprintTitle{%
 \let\@oddhead\@empty
 \let\@evenhead\@empty
 \def\@oddfoot{}%
 \let\@evenfoot\@oddfoot}
\makeatother

\begin{document}

\begin{frontmatter}


\title{Effect of Electric Charge on Biotherapeutic Transport, Binding and Absorption: A Computational Study}



\author{Mario de Lucio$^*$\corref{Corresponding author:mdeluci@purdue.edu}}
\author{Pavlos P. Vlachos}
\author{Hector Gomez}
\cortext[cor1]{Corresponding author: mdeluci@purdue.edu}

\address{School of Mechanical Engineering, Purdue University, 585 Purdue Mall, West Lafayette, Indiana 47907, USA}

\begin{abstract}
This study explores the effects of electric charge on the dynamics of drug transport and absorption in subcutaneous injections of monoclonal antibodies (mAbs). We develop a novel mathematical and computational model, based on the Nernst-Planck equations and porous media flow theory, to investigate the complex interactions between mAbs and charged species in subcutaneous tissue. The model enables us to study short-term
transport dynamics and long-term binding and absorption for two mAbs with different electric properties. We examine the influence of buffer pH, body mass index, injection depth, and formulation concentration on drug distribution and compare our numerical results with experimental data from the literature.
\end{abstract}


\end{frontmatter}


\section{Introduction}
\begin{figure}[h!]
    \centering
    \includegraphics[width=0.9\columnwidth]{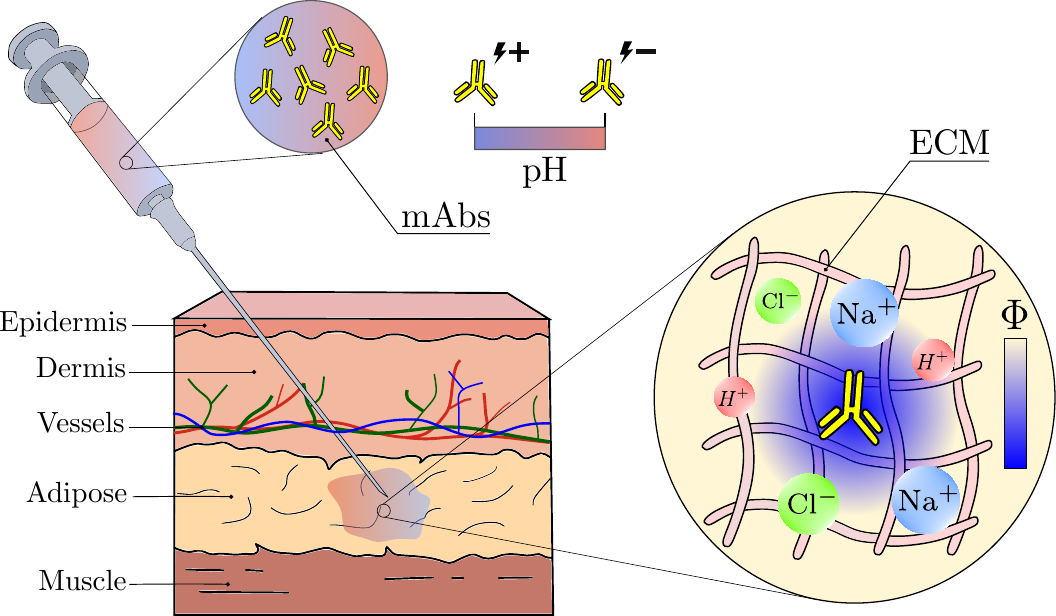}
    \caption{Schematic illustration of the injection process at different scales. Monoclonal antibodies (mAbs) carry an electric charge that varies with the pH level. Upon injection into the adipose layer, mAbs interact with the ion species within the extracellular matrix (ECM) of the tissue, which in turn modifies the local electric potential $\Phi$ and pH. These changes influence the binding rate of mAbs to the ECM.}
    \label{figure1}
\end{figure}

Subcutaneous injections of monoclonal antibodies (mAbs) have revolutionized modern medicine, becoming a cornerstone in the treatment of various diseases, including cancer and autoimmune disorders. The efficiency of these therapies is governed by complex mechanical, chemical, and electric interactions regulating drug biotransport  and bioabsorption in subcutaneous tissues. Fig.~\ref{figure1} shows a schematic illustration of drug administration via subcutaneous injection, where mAbs are delivered into the interstitial area underneath the dermis, commonly referred to as the adipose layer. The interstitium is composed of an extracellular matrix (ECM) made up of collagen, negatively charged glycosaminoglycans (GAGs), and ions \cite{lymphflow1,lymphflow2}. The ionic composition of the interstitial fluid (ISF) is presented in Table \ref{table1}, with sodium Na$^+$ and chloride Cl$^-$ comprising 91\% of the total ions. mAbs possess ionizable acidic and basic groups, which can lead to a net positive or negative charge depending on the surrounding environment \cite{mAbscharge1}. This charge influences their interstitial transport through electrostatic interactions with the negatively charged ECM and the ionic species, thereby impacting binding, absorption, and bioavailability \cite{MCLENNAN200589}. 

Positively charged mAbs exhibit increased binding to the ECM, potentially reducing absorption \cite{mAbelectric}. Positive charge also promotes cellular uptake, and increased blood clearance \cite{liu2021effect,boswell2010effects}. The buffer pH used in mAb formulations plays a crucial role in determining the net charge  and binding kinetics of the antibodies, which directly influences their interaction with the ECM and ionic species in the interstitial space \cite{buffer1}. When the buffer solution is injected, it may alter the local tissue pH, which in turn affects the charge of the mAb and its binding kinetics. At lower buffer pH, mAbs may become more positively charged, leading to enhanced binding to the ECM but potentially reducing systemic absorption due to increased tissue retention. Conversely, higher buffer pH can result in a more negatively charged mAb, reducing ECM binding and improving transport and absorption \cite{mAb1_exps,mAb2_exps}. Thus, studying the effect of buffer pH is crucial for understanding local tissue interactions and drug distribution.

Pharmacokinetic/pharmacodynamic (PKPD) models have been widely used to describe mAb distribution and clearance, and some models account for the effect of charge on drug absorption \cite{liu2023physiologically,patidar2024minimal}. However, these models generally do not capture spatial variations in the tissue, as they are based on ordinary differential equations (ODEs). On the other hand, biomechanical models based on partial differential equations (PDEs) have been developed to incorporate both spatial and temporal distributions of mAbs in tissue, allowing for a more accurate representation of drug transport and absorption. Many of these models are based on a coupling of a poroelastic material framework \cite{DELUCIO1,Leng3,DELUCIO2026126431} with drug transport dynamics \cite{DeLucioThesis,DELUCIO2024124446}, and some extend this approach to include more complex constitutive material models, such as porohypereslatic \cite{LENG2021113919, DELUCIO2023105602,deLucio3}, or even poroviscoelastic \cite{LENG2,Jacques1}. Additionally, other researchers have utilized multicompartment models, which explicitly account for fluid exchanges between the blood, interstitium, and lymphatic vessels, offering a more comprehensive understanding of mAb transport and distribution \cite{haowang1,WANG2023116362,hao3}. However, these models often omit the role of charge, focusing primarily on fluid flow, diffusion, and tissue mechanics.

In this work, we develop a multiphysics computational model to investigate the effect of electric charge on mAb transport, binding, and absorption during subcutaneous injection. The model incorporates three main mechanisms: advection, driven by porous media flow, diffusion and electromigration. We account for the pH-dependency of the drug charge and binding kinetics. We apply this model to study transport and binding phenomena across various buffer pH levels, body mass indices and injection depths.

\begin{table}[h]
\begin{center}
\caption{Ionic composition of Interstitial Fluid (ISF). Extracted from \cite{CSF}.}
\label{Cases}
\begin{tabular}{ll}
\hline
Species & ISF \\
\hline
Na$^+$   & 146 mM \\
K$^+$    & 4.1 mM\\
Cl$^-$     & 118 mM\\
HCO$_3^-$  & 22.0 mM\\
H$^+$      &pH=7.44 
\\\hline
\end{tabular}
\label{table1}
\vspace*{-4pt}
\end{center}
\end{table}

\section{Governing equations}
\label{S:1}
The core of our computational framework lies in the Nernst-Planck equations \cite{NP,BAUER2012199}. These equations describe the diffusion and electromigration of ionic species within a medium, considering both the influence of concentration gradients and the electric potential. In our model, the Nernst-Planck equations are coupled with a porous media flow model based on Darcy's law. This set of equations provides the framework to comprehend the interplay between charged species during subcutaneous injections of mAbs. In this model, the mass conservation of each ion species can be written as
\begin{equation}
\frac{\partial \left(c_i n\right)}{\partial t}+\nabla\cdot \bm{j}_i =q_i,
\label{mass_balance}
\end{equation}
where $c_i$ is the molar concentration of each ion species, $n$ is the porosity of the medium, which we assume is constant, $t$ is time, $q_i$ is a reactive sink/source term, and $\bm{j}_i$ is the mass flux vector
\begin{equation}
    \bm{j}_i=\underbrace{\bm{u}c_i}_{\text{Advection}}-\underbrace{D_i n \nabla c_i}_{\text{Diffusion}} -\underbrace{z_i\frac{D_i F}{R T}n c_i \nabla \Phi}_{\text{Electromigration}} \qquad \text{ (no sum on $i$) }.
\end{equation}
Here, $\bm{u}$ is the fluid velocity, $D_i$ is the diffusion coefficient of the $i$-th species, $z_i$ is the ionic valence, $F$ is the Faraday constant, $R$ is the universal gas constant, $T$ is the temperature, and $\Phi$ is the electric potential. We model fluid flow using Darcy’s law, and assuming that the fluid is incompressible,
\begin{equation}
\bm{u}=-\frac{\kappa}{\eta}\nabla p, \text{ with } \nabla \cdot \bm{u}=q_p+J_b-J_l,
\end{equation}
where $\kappa$ is the permeability of the porous medium, $\eta$ is the fluid dynamic viscosity, $p$ is the pore pressure, $q_p$ is a source term representing the injection of solvent or buffer solution, and $J_b$, $J_l$ model the blood filtration and lymphatic uptake, respectively. The electric potential $\Phi$ is an additional unknown scalar field in Eqn.~\eqref{mass_balance}, which is determined from the electroneutrality condition
\begin{equation}
\sum_{i=1}^m z_i c_i =0.
\label{electroneutrality}
\end{equation}
This condition is an algebraic constraint originating from the assumption that the electrolyte solution is locally electrically neutral. In a macroscopic model, such as the one we are interested in, Eqn.~\eqref{electroneutrality} provides an adequate and generally accepted approximation. By using the electroneutrality condition, we can eliminate one ionic concentration from the system of equations using
\begin{equation}
    c_m=-\frac{1}{z_m}\sum_{i=1}^{m-1} z_ic_i,
\end{equation}
and assuming $z_m \neq 0$. If we now multiply each transport equation by $z_i$, sum over $i$, and assume that the $z_{i}$'s are constant, we obtain the equation
\begin{equation}
    -\nabla \cdot \left[\sum_{i=1}^{m-1}z_i n \left(D_i-D_m\right)\nabla c_i\right]-\nabla \cdot \left[\left(\sum_{i=1}^{m-1}z_iF n\left(z_i \mu_i-z_m\mu_m\right)c_i \right)\nabla \Phi \right]=\sum_{i=1}^{m} z_iq_i,
    \label{electro2}
\end{equation}
where $\mu_i=D_i/R/T$ is the mobility of the ion species.

In our study, the ISF and mAbs are characterized by four distinct ion species and, thus, $m=4$. Given that the ISF predominantly consists of Na$^+$ and Cl$^-$ ions (refer to Table \ref{table1}), we model the ISF using these two ion species. The monoclonal antibodies are modeled as a single pseudo ionic species, whose net charge depends on pH. To be able to model the electric charge of the mAb molecules, our model incorporates an equation dedicated to monitoring the concentration of the hydrogen ion H$^+$, facilitating the calculation of the tissue pH as $\text{pH}=-\log_{10}{\mathrm{H}^+}$. Using the notation $c_1=c_{\text{Na}^+}, c_2=c_{\text{H}^+},c_3=c_{\text{mAb}^+}$ and $c_4=c_{\text{Cl}^-}$, the system of equations reads
\begin{align}
\footnotesize
&\frac{\partial \left(c_{\text{Na}^+} n\right)}{\partial t}+\nabla\cdot \left({\bm{u}c_{\text{Na}^+}}-D_{\text{Na}^+} n \nabla c_{\text{Na}^+} -z_{\text{Na}^+}F\mu_{\text{Na}^+} n c_{\text{Na}^+} \nabla \Phi\right)= q_{p} c_{\text{Na}^+}^{\text{max}}, \label{con_Na}\\
 &\frac{\partial \left(c_{H^+} n\right)}{\partial t}+\nabla\cdot \left({\bm{u}c_{H^+}}-D_{{H^+}} n \nabla c_{H^+} -z_{H^+}F\mu_{\text{H}^+}n c_{{H^+}} \nabla \Phi\right)=q_{p} c_{\text{H}^+}^{\text{max}}, \label{con_H}\\
&\frac{\partial \left(c_{\text{mAb}} n\right)}{\partial t}+\nabla\cdot \left({\bm{u}c_{\text{mAb}}}-D_{\text{mAb}} n \nabla c_{\text{mAb}} -z_{\text{mAb}}\left(\text{pH}\right)F\mu_{\text{mAb}}n c_{\text{mAb}} \nabla \Phi\right)=  \label{con_mAb}\\
& \notag q_{p} c_{\text{mAb}}^{\text{max}} -J_l c_{\text{mAb}}-\phi_B, \\
& \nabla \cdot \left[\sum_{i=1}^{3}z_i n \left(D_i-D_{\text{Cl}^-}\right)\nabla c_i\right]+\nabla \cdot \left[\left(\sum_{i=1}^{3}z_iF n \left(z_i\mu_i-z_{\text{Cl}^-}\mu_{\text{Cl}^-}\right)c_i \right)\nabla \Phi \right]= \label{electroneutrality2} \\
& \notag z_{\text{mAb}}\left(\text{pH}\right)\left(J_l c_{\text{mAb}} + \phi_B \right),  \\
&\nabla \cdot \left(-\frac{\kappa}{\eta}\nabla p\right)=q_p+J_b-J_l, \label{darcy} \\
&\frac{\partial c_B}{\partial t}=k_a\left(\text{pH}\right) n               c_{\text{mAb}}\left(B_{\text{max}}-c_B\right)-k_d\left(\text{pH}\right) c_B-k_e c_B. \label{binding}
\end{align}
Note that the concentration of $\text{Cl}^-$ has been removed from the system of the equations, as it is now determined by the algebraic relation $c_{\text{Cl}^-}=-\frac{1}{z_{\text{Cl}^-}}\sum_{i=1}^{m-1} z_ic_i$. However, the influence of Cl$^-$ remains in the formulation through the constraint  imposed by Eqn.~\eqref{electro2}. In Eqns.~\eqref{con_Na} -- \eqref{con_mAb}, $c_{\text{Na}^+}^{\text{max}}$, $c_{\text{H}^+}^{\text{max}}$, $c_{\text{mAb}}^{\text{max}}$ are the concentration of $\text{Na}^+$, $\text{H}^+$ and mAbs in the syringe expressed in mol/volume of fluid. These allow us to control the ionic buffer composition and pH. We assume that the injectate is in electrical equilibrium, thus, $\sum_{i=1}^m z_i c_i^{\text{max}}=0$. We set the concentrations of $\text{Na}^+$ and $\text{Cl}^-$ in the syringe to be three times their physiological levels to mimic the elevated ionic strength commonly used in mAb formulations. This higher ionic concentration helps modulate protein-protein interactions, stabilize the mAb solution, and create electrochemical gradients that enhance transport and absorption processes during subcutaneous injection \cite{Jain2017}. 

The terms $J_b$, $J_l$ represent the blood filtration and lymphatic uptake. We model them using Starling's law, thus,
\begin{equation}
    J_b=n L_{pb}\frac{S_b}{V}\left(p_b-p-\sigma_r\left(\pi_b-\pi_i\right)\right),
\end{equation}
\begin{equation}
    J_l=n L_{pl}\frac{S_l}{V}\left(p-p_l\right),
\end{equation}
where $L_{pb}$, $L_{pl}$ are the hydraulic conductivity of blood and lymphatic vessels; $S_b/V$, $S_l/V$ represent the specific surface area of blood and lymphatic vessels per volume of the tissue; $p_b$, $p_l$ are the pressure in blood and lymphatic capillaries; $\sigma_r$ is the osmotic reflection coefficient of blood capillaries for the drug, and $\pi_b$, $\pi_i$ are the osmotic pressures in blood vessels and the interstitial space, respectively. 

\begin{figure}[h!]
    \centering
    \includegraphics[width=0.99\columnwidth]{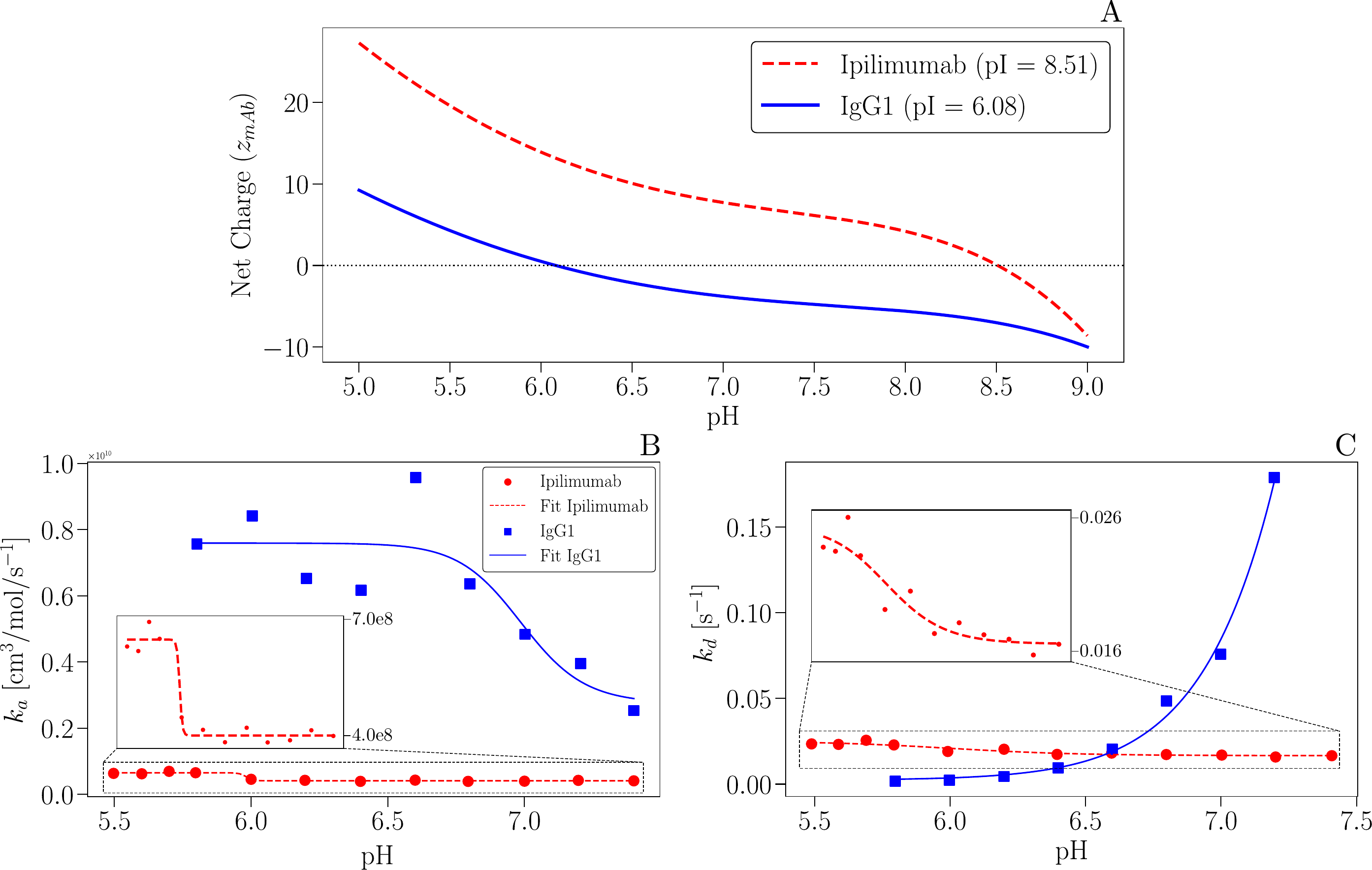}
    \caption{(A) Isoelectric plots of the mAbs considered in the study, where we mark the isoelectric point (pI). The isoelectric points were computed from the amino acid sequences of the mAbs using the \texttt{ProteinAnalysis} module from \texttt{Bio.SeqUtils.ProtParam} in Biopython~\cite{cock2009biopython}. (B-C) Experimental data and fits of the association $k_a$ and dissociation $k_d$ rate at different pH levels \cite{mAb1_exps,mAb2_exps}.}
    \label{e_vs_PH}
\end{figure}

The binding of drug proteins to the ECM is based on a continuum assumption modeled by Eqn.~\eqref{binding}, where $c_B$ is the concentration of mAb bound to the matrix, $k_a$ is the association rate, $B_{\text{max}}$ is the bound concentration at saturation, $k_d$ is the dissociation rate, and $k_e$ is the elimination rate constant. The binding sink term $\phi_B$ in Eqn.~\eqref{con_mAb} reads,
\begin{equation}
    \phi_B=-k_a\left(\text{pH}\right) n c_{\text{mAb}}\left(B_{\text{max}}+c_B\right)+k_d\left(\text{pH}\right) c_B +k_e c_B.
    \label{binding_sink}
\end{equation}
Note that the drug charge, $z_{\text{mAb}}$, and the binding kinetic parameters, $k_a$ and $k_d$, depend on pH. The pH dependence of $z_{\text{mAb}}$ violates the assumption of constant $z_i$ values, meaning a complete derivation of Eqs.~\eqref{con_Na}--\eqref{binding_sink} would include additional terms arising from the variation of $z_{\text{mAb}}$. However, we neglect these derivatives and focus on the first-order effects of the variation of $z_{\text{mAb}}$ with respect to the pH. The net charge of the mAbs changes with pH due to ionization, which in turn influences how the drug interacts with its binding sites. To account for these effects, we incorporated experimental data from the literature, specifically isoelectric plots that relate drug charge to pH and measured relationships between pH, $k_a$, and $k_d$ for two different mAbs. Fig.~\ref{e_vs_PH} shows these data for two molecules of interest, Ipilimumab and IgG1.

The isoeletric plots shown in Fig.~\ref{e_vs_PH}A, provide a clear visualization of how the charge of each mAb changes with pH. At lower pH values, both mAbs tend to be more protonated, resulting in a higher positive charge. As the pH increases the deprotonation occurs, leading to a decrease in net charge. Ipilimumab's isoelectric point (pI) is higher than that of IgG1, leading to a higher positive charge for low pH values. In Figs.~\ref{e_vs_PH}B and C, we fit curves for the experimental data of $k_a$ and $k_d$ as functions of pH. These fits show that lower pH leads to higher association rates for both mAbs, while higher pH values result in a lower dissociation rate for Ipilimumab and a higher one for IgG1. For Ipilimumab, the data shown correspond specifically to the unmodified Ipilimumab (Ipi) antibody reported in \cite{mAb1_exps}, and not to any of the engineered or variant forms included in that study. For IgG1, the data are taken from \cite{mAb2_exps} and specifically correspond to the human IgG1 mAb1 YTE variant evaluated for FcRn binding across pH 5.8–7.4.

\section{Numerical method}
The governing Eqns.~\eqref{con_Na} -- \eqref{binding_sink} were implemented in the open-source finite element library \texttt{FEniCS} \cite{fenics}. The spatial discretization of the governing equations was performed using linear finite elements. We use a staggered time integration scheme, where we first solve Eqns.~\eqref{electroneutrality2} and \eqref{darcy} and to obtain the electric potential $\Phi$ and the fluid velocity $\bm{u}$. These values are then substituted into the advection-diffusion-electromigration equations, Eqns.~\eqref{con_Na} -- \eqref{con_mAb}, and used to compute the concentrations of Na$^+$, Cl$^-$, H$^+$, and mAbs. Finally we update the binding kinetics parameters, $k_a$ and $k_d$ with the pH and solve the binding equation, Eqn.~\eqref{binding}. For each of the substeps in the staggering scheme, we discretize in time using backward Euler. The resulting linear systems of equations are solved using the MUMPS direct solver.

\subsection{Computational setup}
Assuming axisymmetry, we model the injection process in a cylindrical domain of tissue with radius $r=5$ cm and height $z=5$ cm. The skin surface is located at $z=5$ cm. The injection site is on the symmetry axis, $r=0$ cm. The source term $q_p$, representing the injection, is modeled using the multiplicative decomposition proposed in \cite{DELUCIO1}. Unless otherwise specified, we consider an injection depth of 8 mm and a flow rate of 1 mL over 5 seconds. These values are representative of a subcutaneous injection of mAbs \cite{flowrate1,flowrate2,flowrate3}.  

We apply flux-free boundary conditions for the electric potential. For the fluid pressure, we use flux-free boundary conditions in all boundaries, except for the right boundary ($r=5$ cm), where we impose $p=0$ to simulate an non-impermeable boundary. Similarly, we apply flux-free boundary conditions for all ionic species, ensuring no net movement of ions across the boundaries. We employ a nonuniform mesh refined around the injection site to resolve the large pressure gradients produced by the injection.

\section{Model parameters}
The values of the model parameters are listed in Tables \ref{parameters} and \ref{parameters_LU}. We consider a three-layer porous medium domain consisting of dermis-epidermis, adipose and muscle layers. Each layer has different permeability and surface area of lymphatics per unit volume. We use average thicknesses of 2 mm for the dermis and epidermis, 15 mm for the adipose tissue, and 33 mm for the muscle layer \cite{thickness1,thickness2,thickness3}. The permeability of the tissue varies considerably across the different layers \cite{REDDY1981879,guyton}. Reported values for the adipose layer range between $10^{-11}$ and $10^{-8}$ cm$^2$ \cite{kim2017effective,Thomsen,RAHIMI2022104228}, while the dermis and epidermis exhibit much lower permeability, with values between $10^{-14}$ and $10^{-12}$ cm$^2$ \cite{shrestha2020imaging}. This indicates that the dermis and epidermis are nearly impermeable compared to the adipose layer. Preclinical studies also suggest that the fascia, a dense connective tissue layer that separates and supports muscles and other organs, forms a nearly impenetrable barrier to fluid injected subcutaneously at the adipose-muscle interface \cite{RAHIMI2}. We use permeability values of $10^{-10}$ cm$^2$ for the dermis-epidermis layer, $10^{-9}$ cm$^2$ for the adipose and $10^{-11}$ cm$^2$ for the muscle. We approximate the viscosity of the interstitial fluid using the viscosity of water \cite{DELUCIO1}. We assume a porosity value of $n=0.1$ \cite{Sachs1,Nakagawa}.

The initial ion concentrations in the tissue are based on physiological levels \cite{CSF}. For Na$^+$ and Cl$^-$, the initial concentration in the ECM is 140 mM. The initial concentration of H$^+$ is determined by the acidic pH level of human tissue, where pH = 7.4. We assume no drug (mAbs) is present in the tissue at the initial state. The diffusion coefficients of the ions are based on their values in water \cite{robinson1959electrolyte}. The diffusion coefficient of mAbs in the tissue is assumed to be $10^{-6}$ \cite{HAN2021,RAHIMI2022104228}.

\begin{table}[h!]
\caption{Values of the transport and flow parameters}
\centering
\begin{tabularx}{\textwidth}{X l}
\toprule
\textbf{Parameter} & \textbf{Value} \\
\midrule
Diffusion coeff. Na\(^+\) ($D_{\mathrm{Na}^+}$) & $1.33\times 10^{-5}$ (cm\(^2\)/s) \\
Diffusion coeff. Cl\(^-\) ($D_{\mathrm{Cl}^-}$) & $2.03\times 10^{-5}$ (cm\(^2\)/s) \\
Diffusion coeff. mAbs ($D_{\mathrm{mAb}}$) & $10^{-6}$ (cm\(^2\)/s) \\
Diffusion coeff. H\(^+\) ($D_{\mathrm{H}^+}$) & $9.31\times 10^{-5}$ (cm\(^2\)/s) \\
Ionic valence Na\(^+\) ($z_{\mathrm{Na}^+}$) & $+1$ \\
Ionic valence Cl\(^-\) ($z_{\mathrm{Cl}^-}$) & $-1$ \\
Effective ionic valence mAbs ($z_{\mathrm{mAb}}$) & Variable (see Fig.~\ref{e_vs_PH}A) \\
Ionic valence H\(^+\) ($z_{\mathrm{H}^+}$) & $+1$ \\
Initial concentration Na\(^+\) & $1.4  \times 10^{-4}$ (mol/cm\(^3\)) \\
Initial concentration Cl\(^-\) & Computed using electroneutrality \\
Initial concentration mAbs & 0 \\
Initial concentration H\(^+\) & $4.0\times 10^{-11}$ (mol/cm\(^3\)) for a pH=7.4 (acidic) \\
Faraday Constant ($F$) & 96485 (C/mol) \\
Temperature ($T$) & 293 (K) \\
Universal gas constant ($R$) & 8.314 (J/K/mol) \\
Permeability dermis-epidermis ($\kappa_{\text{D-E}}$) & $10^{-10}$ (cm$^2$) \\
Permeability adipose ($\kappa_{\text{adipose}}$) & $10^{-9}$ (cm$^2$) \\
Permeability muscle ($\kappa_{\text{muscle}}$) & $10^{-11}$ (cm$^2$) \\
Viscosity ($\mu$) & $10^{-7}$ (N$\cdot$s/cm$^2$) \\
Porosity ($n$) & $0.1$ \\
\bottomrule
\label{parameters}
\end{tabularx}
\end{table}

We adopt the blood capillary filtration, lymphatic uptake, and binding parameters as detailed in \cite{HAN2021,RAHIMI2022104228,DeLucioThesis}. Following the approach proposed in \cite{DELUCIO2024124446,RAHIMI2024109193}, we account for the varying presence of lymphatic vessels across different tissue layers. Some studies indicate a higher prevalence of initial lymphatics in the dermal layer compared to the muscle layer  \cite{lymph5,lymph6}, and a greater presence in the adipose layer relative to the muscle layer \cite{lymph7}. However, other research reports the absence of initial lymphatics in the adipose tissue \cite{lymph8}. Due to the lack of experimental data, and to the best of our knowledge, we express the surface area of lymphatics per unit volume as a percentage relative to the initial lymphatics in the dermis-epidermis layer. For the dermis-epidermis layer, we take a value of $S_l/V_{\text{D-E}}=70$ cm$^{-1}$ \cite{HAN2021,RAHIMI2022104228}. We assume that the surface area of lymphatic vessels in the adipose layer is 5\% of $S_l/V_{\text{D-E}}$, and there are no initial lymphatics in the muscle layer \cite{DELUCIO2024124446, RAHIMI2024109193}.
\begin{table}[h!]
\caption{Values of the blood uptake, lymphatic uptake, and binding parameters}
\centering
\begin{tabularx}{\textwidth}{X l}
\toprule
\textbf{Parameter} & \textbf{Value} \\
\midrule
Hydraulic cond. blood vessels ($L_{pb}$) & $10^{-6}$ (cm\(^3\)/N/s) \\
Hydraulic cond. lymph. vessels ($L_{pl}$) & $6.0\times 10^{-5}$ (cm\(^3\)/N/s) \\
Surf. area blood vessels ($S_b/V$) & 70 (1/cm) \\
Surf. area lymph. vessels D-E ($S_l/V_{\mathrm{D-E}}$) & 70 (1/cm) \\
Surf. area lymph. vessels adipose & 5\% $S_l/V_{\mathrm{D-E}}$ (1/cm) \\
Surf. area lymph. vessels muscle  & 0 (1/cm) \\
Pressure blood vessels ($p_b$) & 0.35 (N/cm$^2$) \\
Pressure lymph. vessels ($p_l$) &  0 (N/cm$^2$)\\
Reflection coefficient ($\sigma_r$) & 0.3 \\
Osmotic pressure blood vessels ($\pi_b$) & 0.35 (N/cm$^2$) \\
Osmotic pressure insterstitium ($\pi_i$) & 0.15 (N/cm$^2$) \\
Association rate constant ($k_a$) &  Variable (see Fig.~\ref{e_vs_PH}B) \\
Dissociation rate constant ($k_d$) &  Variable (see Fig.~\ref{e_vs_PH}C)  \\
Elimination rate constant ($k_e$) &  0 \\
Max. bound concentration ($B_{\mathrm{max}})$ &  $10^{-9}$ (mol/cm$^3$)\\
\bottomrule
\label{parameters_LU}
\end{tabularx}
\end{table}

\section{Results}
Unless otherwise stated, we assume a drug concentration of 100 mg/mL in the syringe, and a molar mass of 150,000 g/mol. Thus, the molarity would be $c_{\text{mAb}}^{\text{max}}=\frac{100 \times 10^{-3} \text{g/mL}}{150,000 \text{g/mol}}$ (mol/mL).

\subsection{Short-term dynamics of mAb transport}
\begin{figure}[h!]
    \centering
    \includegraphics[width=0.99\columnwidth]{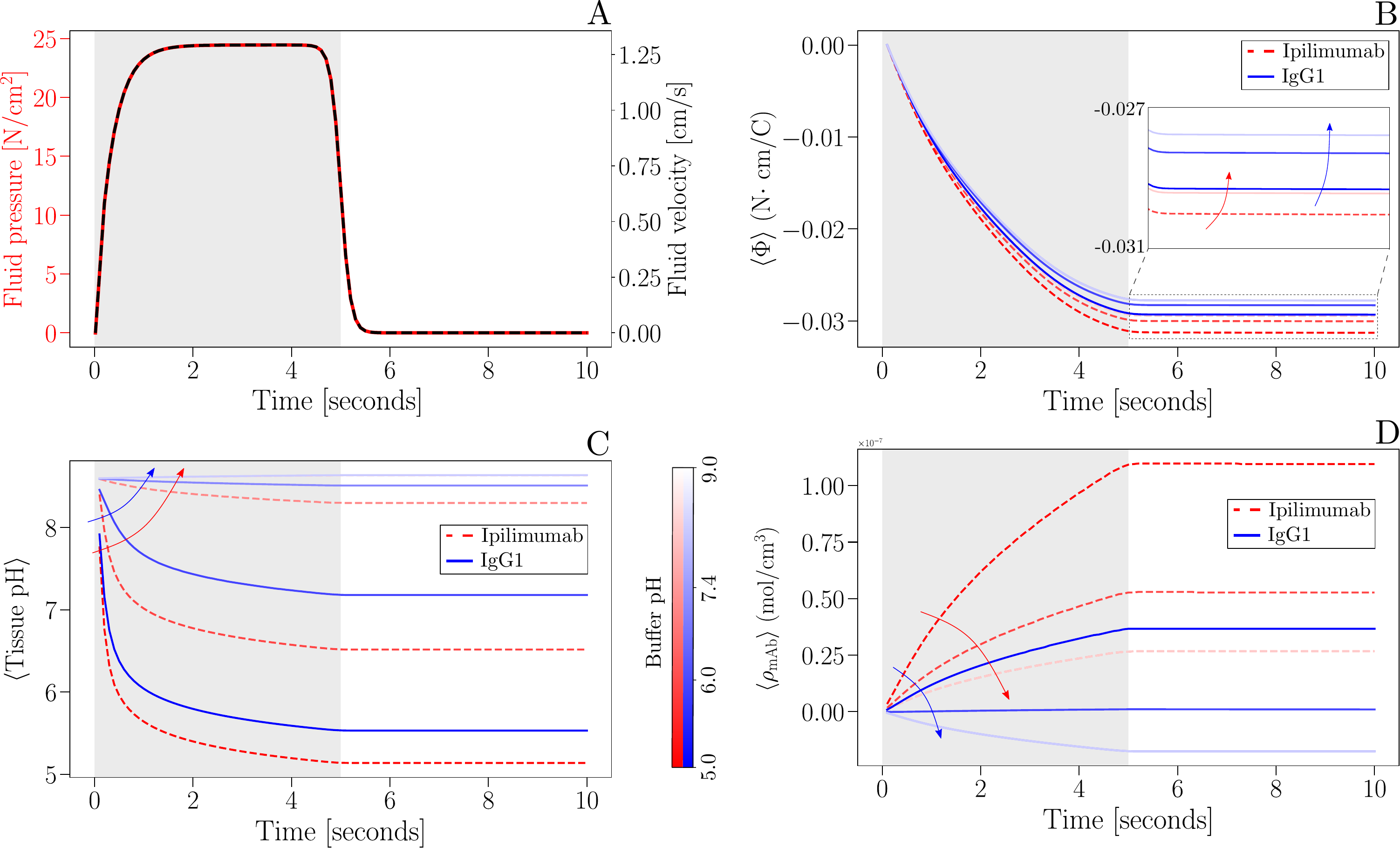}
    \caption{Short-term results for Ipilimumab and IgG1 with different buffer pH levels. (A) Time evolution of fluid pressure and velocity averaged 2 mm around the injection point. (B) Time evolution of averaged electric potential, $\langle \Phi \rangle = 1/|\Omega|\int_{\Omega} \Phi \mathrm{d}\Omega$. (C) Time evolution of averaged tissue pH. (D) Time evolution of the average net charge density, $\langle \rho_{\mathrm{mAb}} \rangle =1/|\Omega|\int_{\Omega} z_{\mathrm{mAb}} c_{\mathrm{mAb}} \mathrm{d}\Omega$. The shaded region denotes the injection time interval. The arrows indicate increasing buffer pH.}
    \label{results}
\end{figure}

Fig.~\ref{results}A illustrates the time evolution of fluid pressure and velocity averaged over a 2 mm radius around the injection point. The injection time interval, $t \in \left[0,5\right]$ is represented by a gray shaded area in the plots. The pressure and velocity profiles closely follow the behavior of the source term $q_p$: they rise rapidly, reach a plateau for rest of the injection time, and drop to near zero after the injection ends. This behavior highlights the significance of advective transport during the injection period. The maximum averaged pressure around the injection site is 25 N/cm$^2$, consistent with values reported in other numerical studies for a rigid porous medium at a flow rate of 720 mL/h \cite{HAN2021}. In our simulation, the maximum velocity around the needle tip is 1.25 cm/s, which closely matches the theoretical velocity of 1.26 cm/s obtained by dividing the volumetric flow rate by the needle area. Thus, we consider our numerical result accurate. 

Figs.~\ref{results}B-D present the domain-averaged time evolution of several variables, computed as $\langle \cdot \rangle = 1/|\Omega|\int_{\Omega} \cdot \ \mathrm{d}\Omega$ for different buffer pH levels for both Ipilimumab and IgG1. The averaged electric potential $\langle \Phi \rangle$ remains negative throughout the injection process regardless of the buffer pH used. It initially decreases exponentially and then remains flat after the injection ends. For a given buffer pH, Ipilimumab consistently shows a lower electric potential compared to IgG1. The tissue pH follows a similar trend, where lower buffer pH results in lower average tissue pH. After the end of the injection, the tissue pH stabilizes just above the injected buffer pH level.  Ipilimumab also shows consistently lower tissue pH than IgG1 at equivalent buffer pH levels.

In Fig.~\ref{results}D, we depict the time evolution of the averaged net charge drug density $\langle \rho_{\mathrm{mAb}} \rangle =1/|\Omega|\int_{\Omega} z_{\mathrm{mAb}} c_{\mathrm{mAb}} \mathrm{d}\Omega$. This term refers to the average charge per unit volume contributed by the drug molecules in the tissue. When the buffer pH is lower than the drug's isoelectric point, the drug carries a positive charge, increasing the charge density. Conversely, at a pH higher than the isoelectric point, the drug carries a negative charge, lowering the net charge density. The average net charge drug density increases as the buffer pH decreases. The highest net charged density is observed for Ipilimumab at a buffer pH of 5.0. At any given buffer pH, Ipilimumab exhibits a higher net charge density than IgG1. Ipilimumab’s higher charge density means that its molecules carry more charge under the same conditions, leading to stronger electrostatic interactions.  The net charge density of IgG1 only becomes negative at buffer pH = 9.  

\begin{figure}[h!]
    \centering
    \includegraphics[width=0.99\columnwidth]{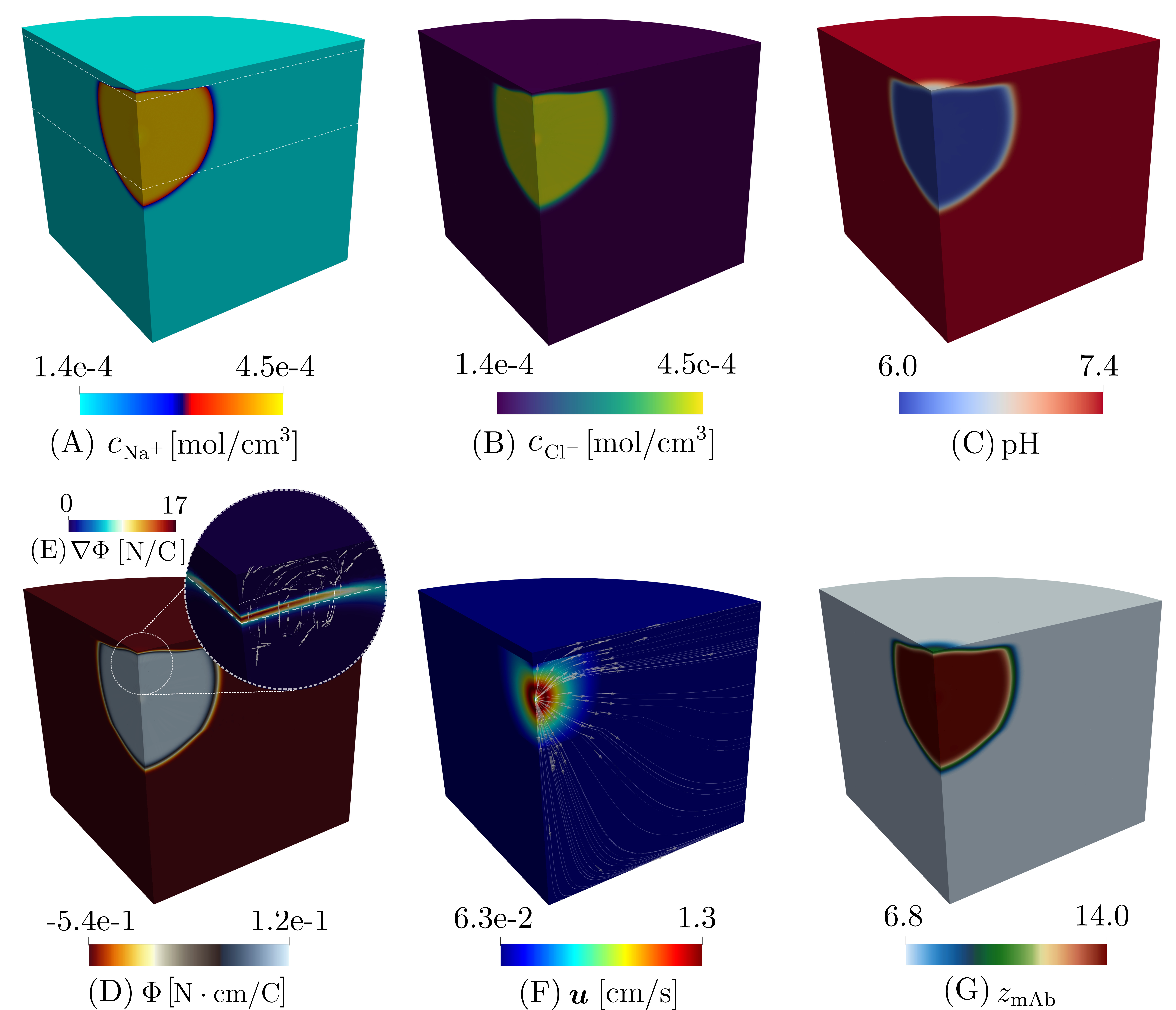}
    \caption{Snapshots of the spatial distribution of different quantities 5 seconds after the end of the injection for the Ipilimumab with a buffer pH of $6.0$. (A) Concentration of Na$^+$. The dashed lines indicate the tissue layer interfaces. (B) Concentration of Cl$^-$. (C) Tissue pH. (D) Electric potential. (E) Close-up of electric potential gradient and streamlines at the edge of the drug depot. (F) Fluid velocity in logarithmic scale and velocity streamlines. (G) Net drug charge.}
    \label{fields}
\end{figure}

Figs.~\ref{fields}A-G show the spatial distribution of several quantities of interest five seconds after the injection of Ipilimumab ends with a buffer pH of 6.0. Since the buffer consists mainly of Na$^+$ and Cl$^-$ ions, their concentrations are nearly equal, as they govern the electroneutrality of the system. The ion concentrations are higher within the drug depot. The pH within the drug depot equals the buffer pH in the syringe. The electric potential inside the drug plume is also lower than in the surrounding tissue. Fig.~\ref{fields}E depicts the gradient of the electric potential and its streamlines. Here, we observe that the edge of the drug depot exhibits a high gradient of electric potential, indicating that electromigration is significant and must be accounted for in the simulations. The velocity field in Fig.~\ref{fields}F is presented on a logarithmic scale to capture the wide range of values. The velocity is highest at the edge of the needle tip and decreases rapidly with distance from it. Finally, Fig.~\ref{fields}G illustrates the spatial distribution of the drug charge. In this case, the charge within the drug depot is higher than outside.

\subsection{Long-term drug binding and absorption dynamics}
We now conduct long-term simulations to evaluate the effect of charge on drug absorption and binding. Given the large number of degrees of freedom and the small time step required for accuracy, performing these simulations for extended periods of time is computationally expensive. To address this, we implement a model reduction strategy. 

As illustrated in Fig.~\ref{results}A, the advective terms become negligible after injection. Thus, we begin by removing Eqn.~\eqref{darcy} from the system, and eliminating all the convective and source terms, which also vanish after the injection ends. Additionally, since the injection is no longer active, the nonuniform mesh refined around the injection site is no longer required. This allows us to project the solution of the final time step of our short-term simulation onto a coarser mesh, enabling us to perform long-term calculations more efficiently. 

\subsubsection{Effect of buffer pH}
First we analyze the effect of buffer pH on the binding and absorption dynamics. Buffer pH is crucial in the subcutaneous delivery of mAbs, as it influences electrostatic interactions between the mAb and the ECM. At certain pH levels, imbalanced charges can lead to aggregation or denaturation, reducing stability and therapeutic efficacy. Electrostatic repulsion or attraction due to pH changes can also alter lymphatic uptake and distribution in tissues \cite{protein_stability_buffer_pH}. 

\begin{figure}[h!]
    \centering
    \includegraphics[width=0.95\columnwidth]{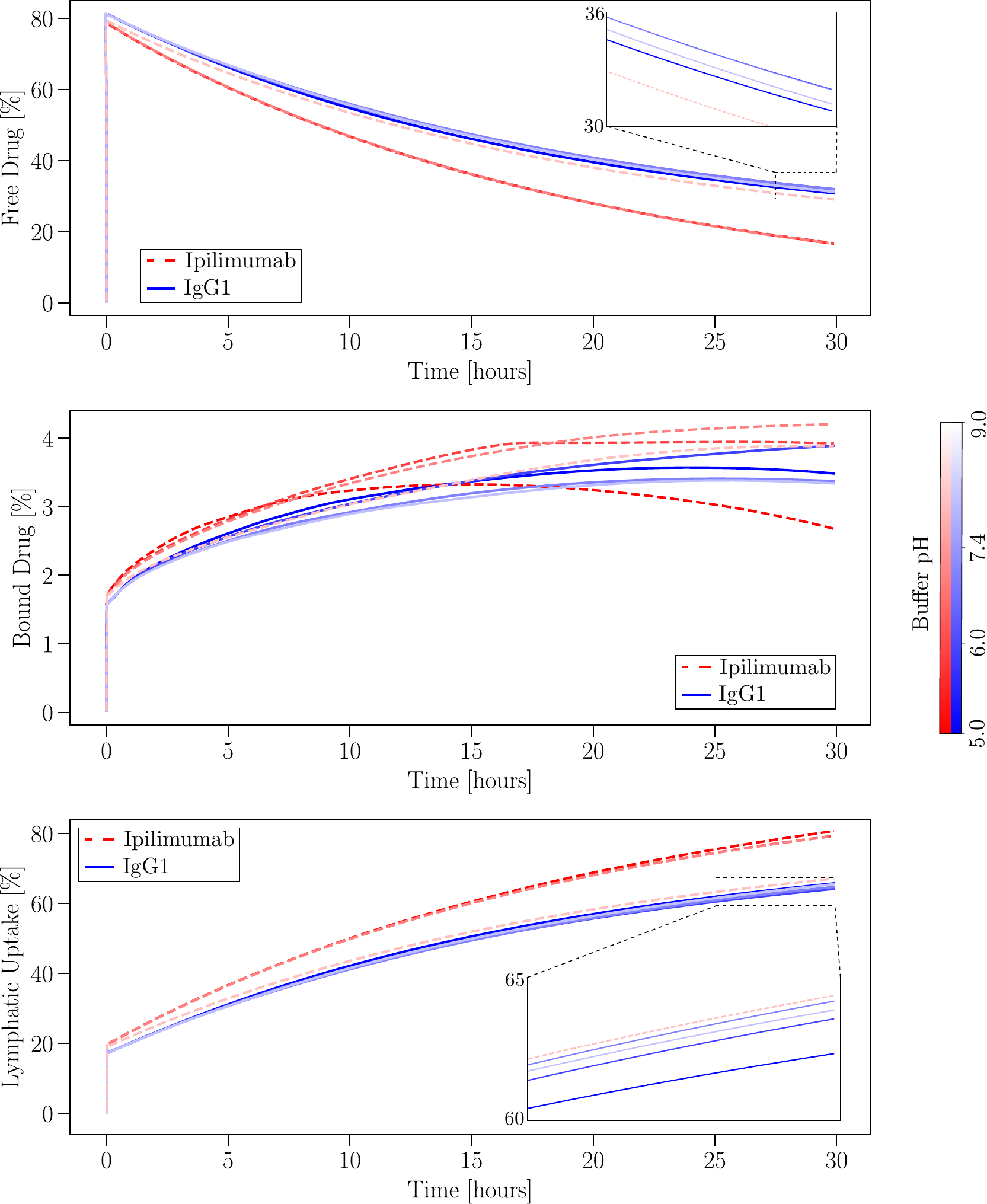}
    \caption{Effect of buffer pH on the long-term distribution of free, bound, and absorbed drug as a percentage of the total injected volume for two different mAbs.}
    \label{buffer_pH}
\end{figure}

Fig.~\ref{buffer_pH} shows the long-term time evolution of the free, bound, and absorbed drug for Ipilimumab and IgG1 at different buffer pH levels. The percentage differences between Ipilimumab and IgG at $t=30$ h are discussed next for free drug, bound drug, and lymphatic uptake levels across buffer pH levels of 5.0, 6.0, 7.4, and 9.0. 

At buffer pH 5.0, Ipilimumab showed significantly lower free drug levels compared to IgG (46\%) and a reduction in bound drug (23\%). However, lymphatic uptake for Ipilimumab was 22\% higher.

At buffer pH 6.0, the free drug levels for Ipilimumab remained lower than IgG1 (47\%), while the bound drug levels were similar (0.87\%), and lymphatic uptake was 23\% higher for Ipilimumab. This indicates that at slightly acidic conditions, Ipilimumab is more easily absorbed through the lymph compared to IgG1.

At buffer pH 7.4, which reflects physiological pH, Ipilimumab continued to exhibit significantly lower free drug levels than IgG1 (47\%). However, the bound drug levels for Ipilimumab increased (24\%), and lymphatic uptake was 22\% higher.

At buffer pH 9.0, the differences between Ipilimumab and IgG1 were less pronounced. The free drug level for Ipilimumab was only 6.8\% lower than IgG1, while bound drug levels were higher by 16\%, and lymphatic uptake was higher by 2.5\%. This results suggests that at alkaline conditions, the pharmacokinetics of Ipilimumab and IgG1 are comparable.

\begin{figure}[h!]
    \centering
    \includegraphics[width=0.99\columnwidth]{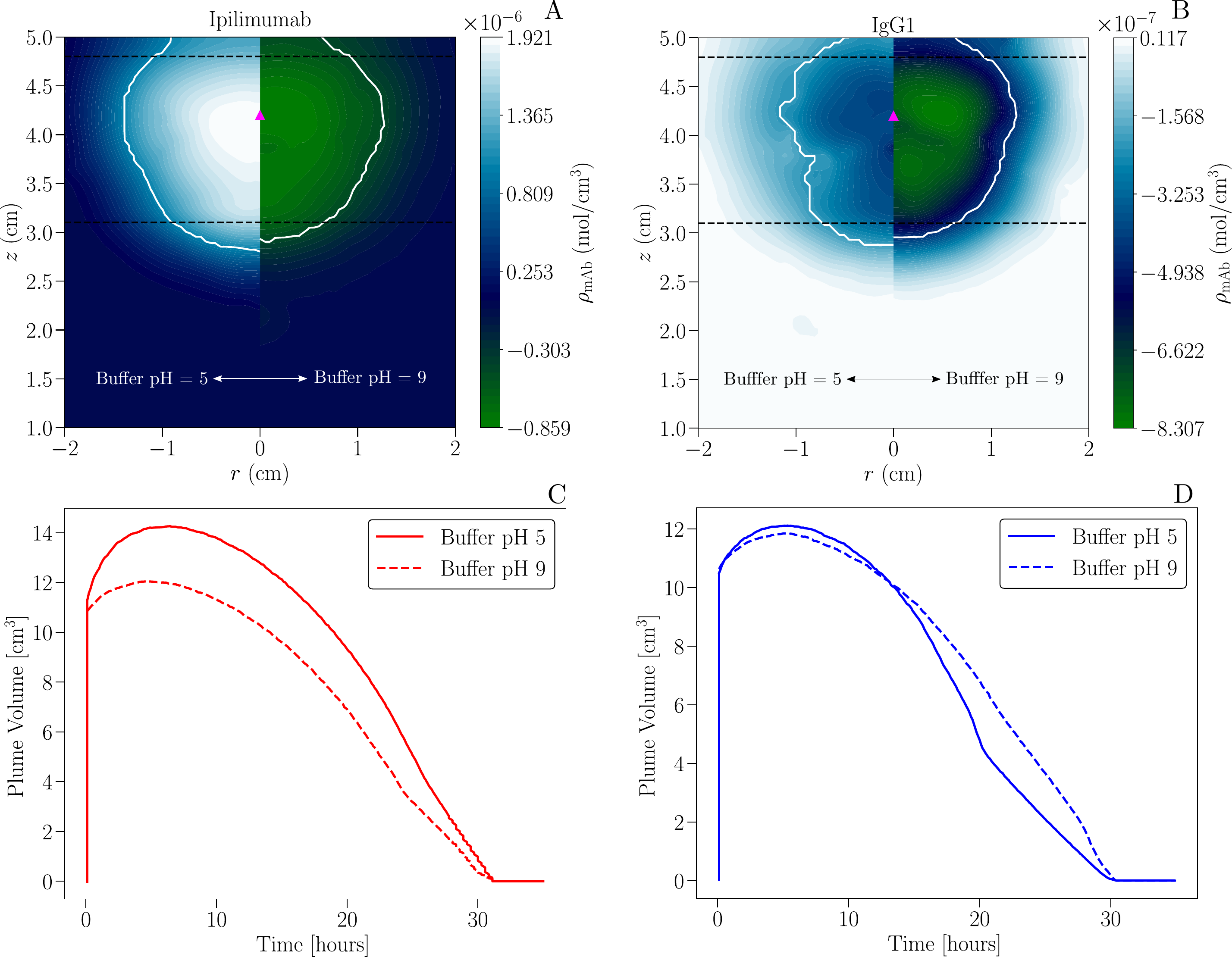}
    \caption{(A-B) Spatial distribution of net charge drug density at $t = 20$ h for Ipilimumab and IgG1 across different buffer pH levels. The white solid line, representing the zero line of $\text{sign} \left( c_{\text{mAb}} - 0.5 \max \left( c_{\text{mAb}} \right) \right)$, delineates the edge of the drug depot. The dashed black lines mark the tissue layer interfaces. The magenta triangle marks the injection point. (C-D) Plume volume over time for different buffer pH levels.}
    \label{buffer_pH2}
\end{figure}

We examine the spatial distribution of net charge drug density, $\rho_{\text{mAb}}$ at $t=20$ h in Figs.~\ref{buffer_pH2}A-B for Ipilimumab and IgG1, and for buffer pH levels of 5 and 9. These plots highlight regions where the drug's net charge is concentrated. For Ipilimumab, at low buffer pH, the drug shows a positive charge density within the drug depot. Conversely, at higher buffer pH, beyond the drug's isoelectric point (pI), the charge density inside the depot becomes negative. In addition, we mark the edge of the drug depot, calculated as the zero line of $\text{sign} \left( c_{\text{mAb}} - 0.5 \max \left( c_{\text{mAb}} \right) \right)$, with a solid white line. We observe that the drug plume is more dispersed at lower buffer pH, likely due to stronger interactions with the negatively charged extracellular matrix. For IgG1 at low buffer pH (below its pI), the drug carries a smaller negative charge (in absolute value) than at a higher buffer pH. Comparing both mAbs, Ipilimumab produces a larger plume than IgG1 at the same buffer pH. However, the maximum absolute charge density in absolute value is eight times higher for IgG1.

Figs.~\ref{buffer_pH2}C-D show the time evolution of the plume volume. In all cases, the plume volume peaks approximately at $t=7$ h, with Ipilimumab exhibiting the greatest plume volume (14 cm\textsuperscript{3} for Ipilimumab vs. 12 cm\textsuperscript{3} for IgG1). For Ipilimumab, the plume grows faster at a buffer pH of 5 compared to pH 9. However, both depots are almost completely absorbed at approximately $t=30$ h. For IgG1, the plume volumes over time are more similar for the two values of buffer pH.

\subsubsection{Effect of body mass index}
To investigate the effect of Body Mass Index (BMI) on long-term binding and absorption dynamics, we simulate cases for high and low BMI by varying the subcutaneous adipose layer thickness, which typically increases with BMI. We use 15 mm adipose layers for cases with high BMI and 6 mm for those with low BMI, based on the reported range in the supplementary material of \cite{subQ-sim}. We use injection depths of $5$ mm and $8$ mm for the low and high BMI cases, respectively. The buffer pH in all the cases studied here is 7.4. 

\begin{figure}[h!]
    \centering
    \includegraphics[width=0.85\columnwidth]{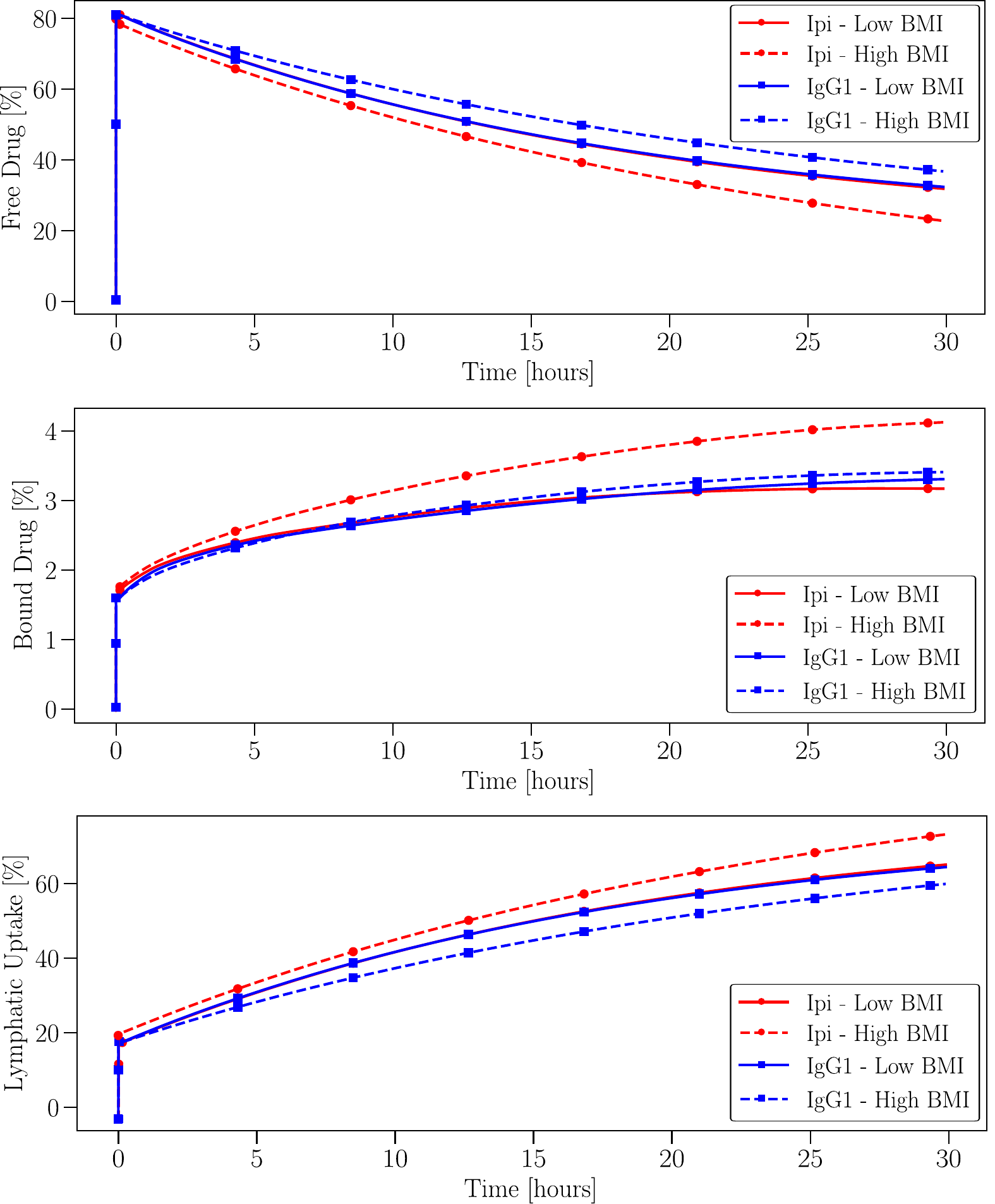}
    \caption{Effect of Body Mass Index (BMI) on the long-term distribution of free, bound, and absorbed drug as a percentage of the total injected volume for two different mAbs.}
    \label{BMI1}
\end{figure}

Fig.~\ref{BMI1} illustrates the time evolution of free, bound, and lymphatic uptake of Ipilimumab and IgG1 for both high and low BMI scenarios. For Ipilimumab, a high BMI significantly reduced free drug levels compared to a low BMI, with a relative difference of 38\%. However, the high BMI case showed a notable increase in bound drug (21\%) and lymphatic uptake (22\%). For IgG1, the impact of BMI was less pronounced, with free drug levels showing a smaller decrease of 11\% and bound drug decreasing by 3\%, while lymphatic uptake increased by 7\%. When comparing Ipilimumab to IgG1 within the same BMI category, Ipilimumab showed slightly lower free drug levels (1.6\%) and bound drug levels (4\%) at low BMI, but a higher lymphatic uptake (1\%). For high BMI, Ipilimumab has lower levels of free drug (38\%) but higher bound drug (21\%) and lymphatic uptake (22\%) compared to IgG1.

\begin{figure}[h!]
    \centering
    \includegraphics[width=0.99\columnwidth]{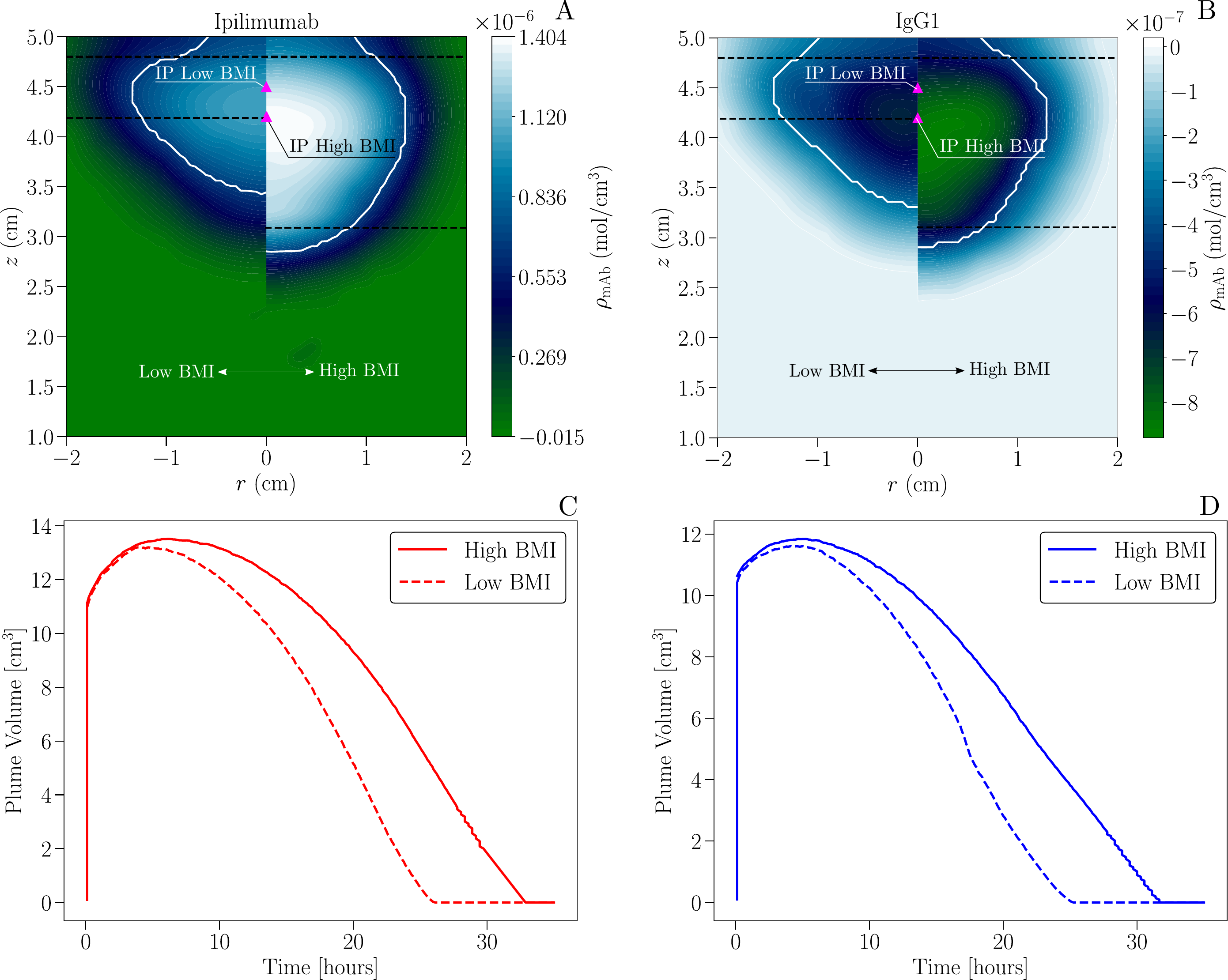}
    \caption{(A-B) Spatial distribution of net charge drug density at $t=20$ h for Ipilimumab and IgG1 across different BMI levels. The white solid line, representing the zero line of $\text{sign} \left( c_{\text{mAb}} - 0.5 \max \left( c_{\text{mAb}} \right) \right)$, delineates the edge of the drug depot. The dashed black lines mark the tissue layer interfaces. The magenta triangle marks the injection point. (C-D) Plume volume over time for different BMI levels.}
    \label{BMI2}
\end{figure}

We analyze the spatial distribution of net charge drug density, $\rho_{\mathrm{mAb}}$, at $t=20$ h in Fig.~\ref{BMI2} for patients with high and low BMI. In both panels, the white solid line delineates the edge of the drug depot, while the dashed black lines mark the interfaces between tissue layers. The pink marker (IP) denotes the injection point. For both mAbs, the distribution is divided into regions representing low BMI (left half of each plot) and high BMI (right half of each plot).

For both mAbs, the charge density in absolute value is higher in the high BMI region compared to the low BMI region. IgG1 exhibits more pronounced differences between the BMI categories. In the low BMI cases, the drug depot shows deeper penetration into the muscle layer, where there is limited lymphatic presence and, thus, minimal drug absorption. Furthermore, the low BMI case reveals that a larger portion of the drug is retained in the dermis-epidermis layer, where most lymphatic vessels are located, which may explain the higher lymphatic uptake observed in low BMI compared to high BMI. In all cases, the maximum net drug charge in absolute value is localized near the injection site within the adipose layer.

We study the time evolution of the plume volume for different BMI levels in Figs.~\ref{BMI2}C-D. The depot edge is defined as the zero line of $\text{sign} \left( c_{\text{mAb}} - 0.5 \max \left( c_{\text{mAb}} \right) \right)$. The peak plume volume at $t \approx 8$ h is similar across the different cases for both mAbs. However, after reaching the maximum, the plume volumes evolve much differently. For both mAbs, the low BMI case exhibits a much faster decrease in plume volume compared to the high BMI case. In both cases, for a low BMI, the plume has disappeared almost completely by $t \approx 25$ h, whereas for a high BMI, it disappears at $t \approx 31$ h.

\subsubsection{Effect of injection depth}
Injection depth is a crucial parameter in subcutaneous injections, as it can vary between devices and even for the same device depending on how it is used by the patient \cite{autoinjector_comparison}. To investigate this factor, we performed simulations with injection depths of 8 mm and 15 mm for both Ipilimumab and IgG1, using a buffer pH of 7.4. While standard subcutaneous injections typically have injection depths from 4 mm to 13 mm \cite{Shi2021}, we also considered an injection depth of 15 mm to investigate the effects of injecting near the muscle layer. This exploratory case allows us to analyze how proximity to muscle tissue influences drug dispersion, absorption, and charge interactions, which is particularly relevant in high-dose biologics where deeper injections may inadvertently occur.

\begin{figure}[h!]
    \centering
    \includegraphics[width=0.85\columnwidth]{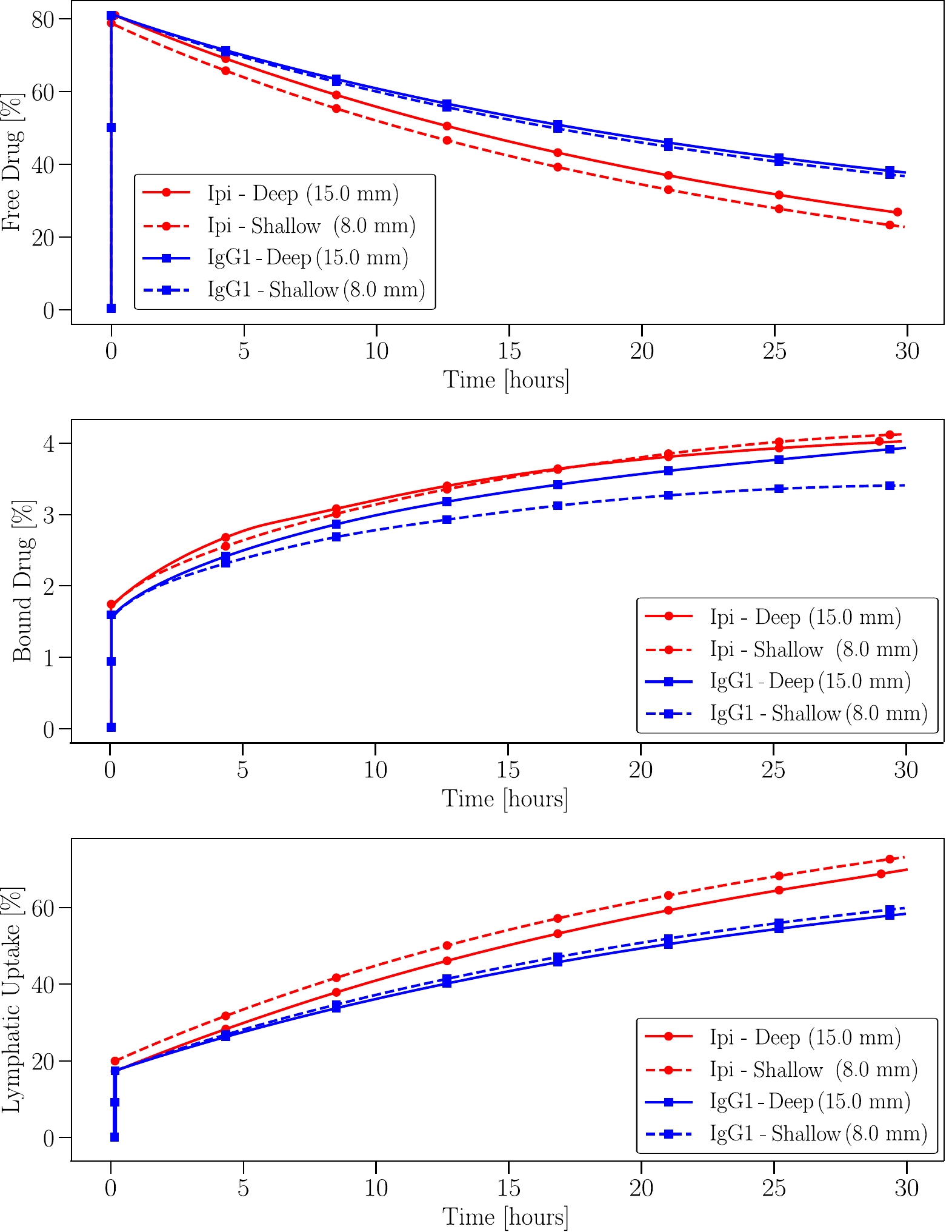}
    \caption{Effect of injection depth on the long-term distribution of free, bound, and absorbed drug as a percentage of the total injected volume for two different mAbs.}
    \label{inj_depth}
\end{figure}

Fig.~\ref{inj_depth} illustrates the time evolution of free, bound, and absorbed drug for Ipilimumab and IgG1 under varying injection depths. For Ipilimumab, a shallower injection reduced free drug levels by 20\% compared to an average injection depth, while bound drug levels increased by 3\%, and lymphatic uptake rose by 8\%. These results suggest that shallower injections improve lymphatic uptake for Ipilimumab.

For IgG1, the impact of injection depth was less pronounced. A shallower injection resulted in a 2.7\% reduction in free drug levels, a 13\% decrease in bound drug, and a modest 2.6\% increase in lymphatic uptake.

\begin{figure}[h!]
    \centering
    \includegraphics[width=0.99\columnwidth]{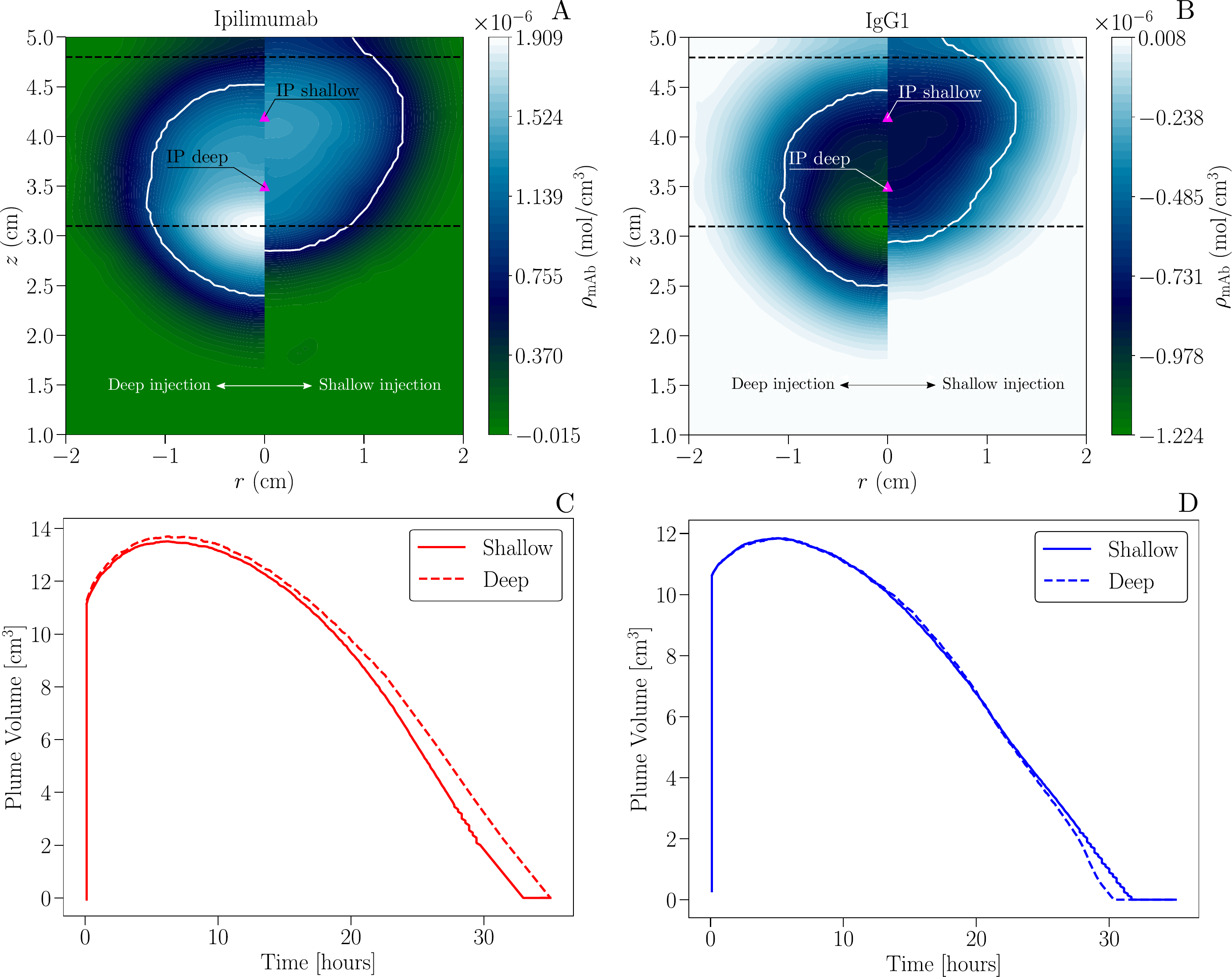}
    \caption{(A-B) Spatial distribution of net charge drug density at $t=20$ h for Ipilimumab and IgG1 across different injection depths. The white solid line, representing the isoline of $\text{sign} \left( c_{\text{mAb}} - 0.5 \max \left( c_{\text{mAb}} \right) \right)$, delineates the edge of the drug depot. The dashed black lines mark the tissue layer interfaces. The magenta triangle marks the injection point. (C-D) Plume volume over time for different injection depths.}
    \label{Charge_Density_depth}
\end{figure}

When comparing Ipilimumab and IgG1 at specific injection depths, the differences become more apparent. For a deep injection, Ipilimumab showed 24.7\% lower levels of free drugs, 1.5\% higher levels of bound drug and 16\% higher lymphatic uptake than IgG1. At a shallower injection depth, the contrast was even more pronounced, with Ipilimumab showing 38\% lower free drug levels, 21\% higher bound drug levels and 22\% greater lymphatic uptake than IgG1. 
Fig.~\ref{Charge_Density_depth}A-B displays the spatial distribution of net charge drug density at $t=20$ h for Ipilimumab and IgG1 across different injection depths. Deep injection scenarios show more penetration of the drug into the muscle layer, with minimal distribution into the dermis-epidermis layer. For these cases, the maximum concentration of net charge drug density is observed at the adipose-muscle interface below the injection site, likely because of the contrast of permeability between the adipose and muscle tissues; the lower permeability of the muscle layer causes drug accumulation at this boundary.
For Ipilimumab, the drug depot for the shallow injection is slightly smaller compared to the deeper injection, reflecting higher drug spreading near the surface. In contrast, the deep injection shows a more localized and concentrated plume near the injection site.

For IgG1, the differences between deep and shallow injections are less pronounced compared to Ipilimumab. While the shallow injection still shows a slightly wider depot, the distribution is relatively uniform for both injection depths.

Figs.~\ref{Charge_Density_depth}C-D depict the plume volume over time for different injection depths. We observe minimal differences between the various injection depths. For Ipilimumab, the plume disappears slightly faster in the shallow injection case. For IgG1, both curves appear nearly identical.

\subsubsection{Effect of formulation concentration}
\begin{figure}[h!]
    \centering
    \includegraphics[width=0.85\columnwidth]{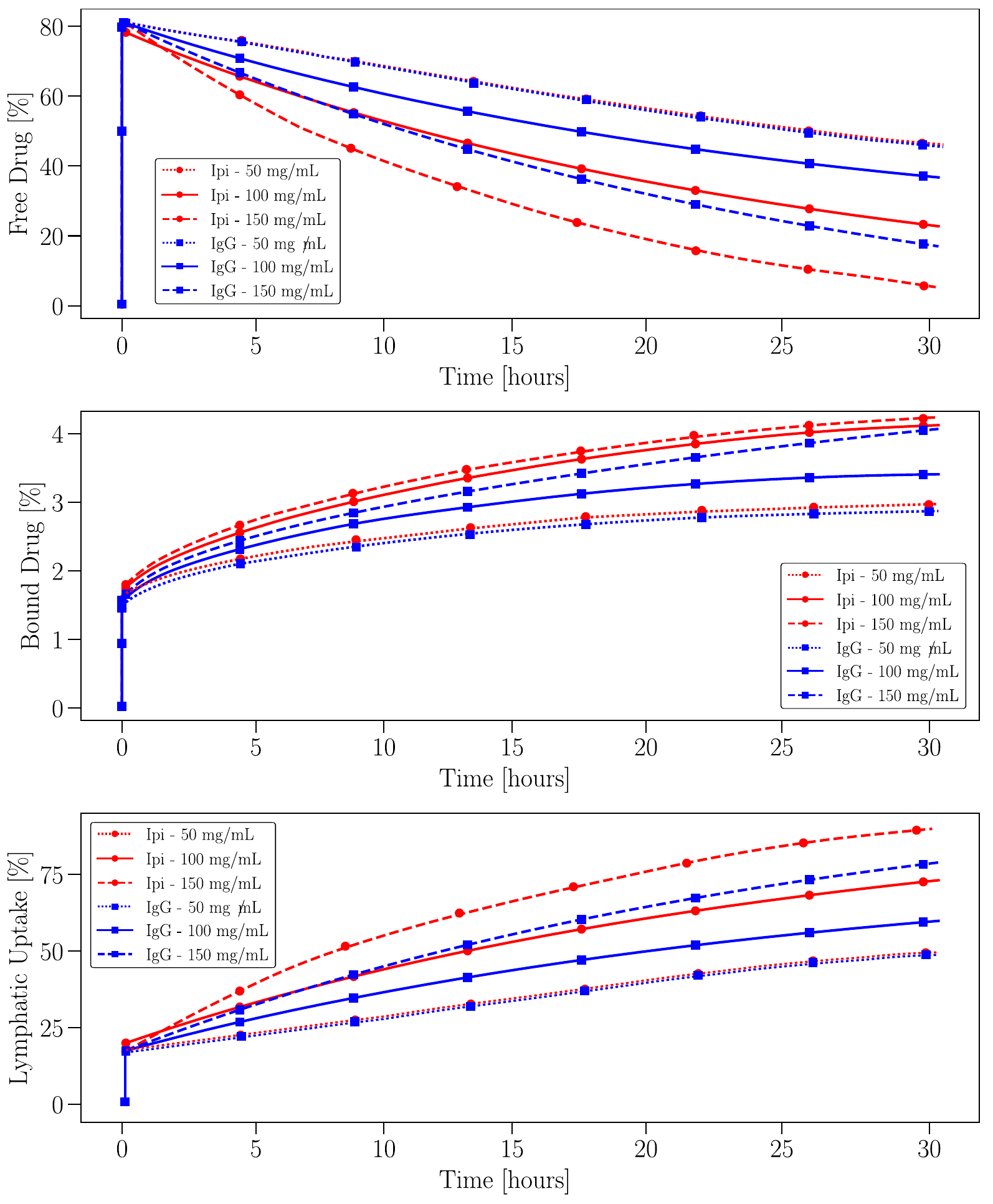}
    \caption{Effect of formulation concentration on the long-term distribution of free, bound, and absorbed drug as a percentage of the total injected volume for two different mAbs.}
    \label{concentration}
\end{figure}

We now investigate the impact of formulation concentration, comparing mAb concentrations of 50, 100 and 150 mg/mL. 
Fig.~\ref{concentration} displays the time evolution of free, bound and absorbed drug. At a lower concentration, the simulations consistently show a higher level of free drug in the tissue, along with a lower bound drug, and lower absorption compared to the higher concentration cases. For both mAbs, the 50 mg/mL profiles exhibit a very similar behavior: the time evolution of free, bound and absorbed drug are almost overlapped. This indicates that the electromigration effects are more pronounced at higher formulation concentrations. Quantitatively, lymphatic uptake increases by approximately 25\% for Ipi and by 15\% for IgG1 when moving from 100 to 150 mg/mL.

From a mechanistic point of view, as the formulation concentration increases, the free drug in the tissue decreases faster because the lymphatic uptake term in Eqn.~\eqref{con_mAb} ($J_l c_{\text{mAb}}$) is directly proportional to the free drug concentration, leading to increased absorption. Meanwhile, binding remains unchanged because the available binding sites either fill up quickly or are already saturated.

\begin{figure}[h!]
    \centering
    \includegraphics[width=0.99\columnwidth]{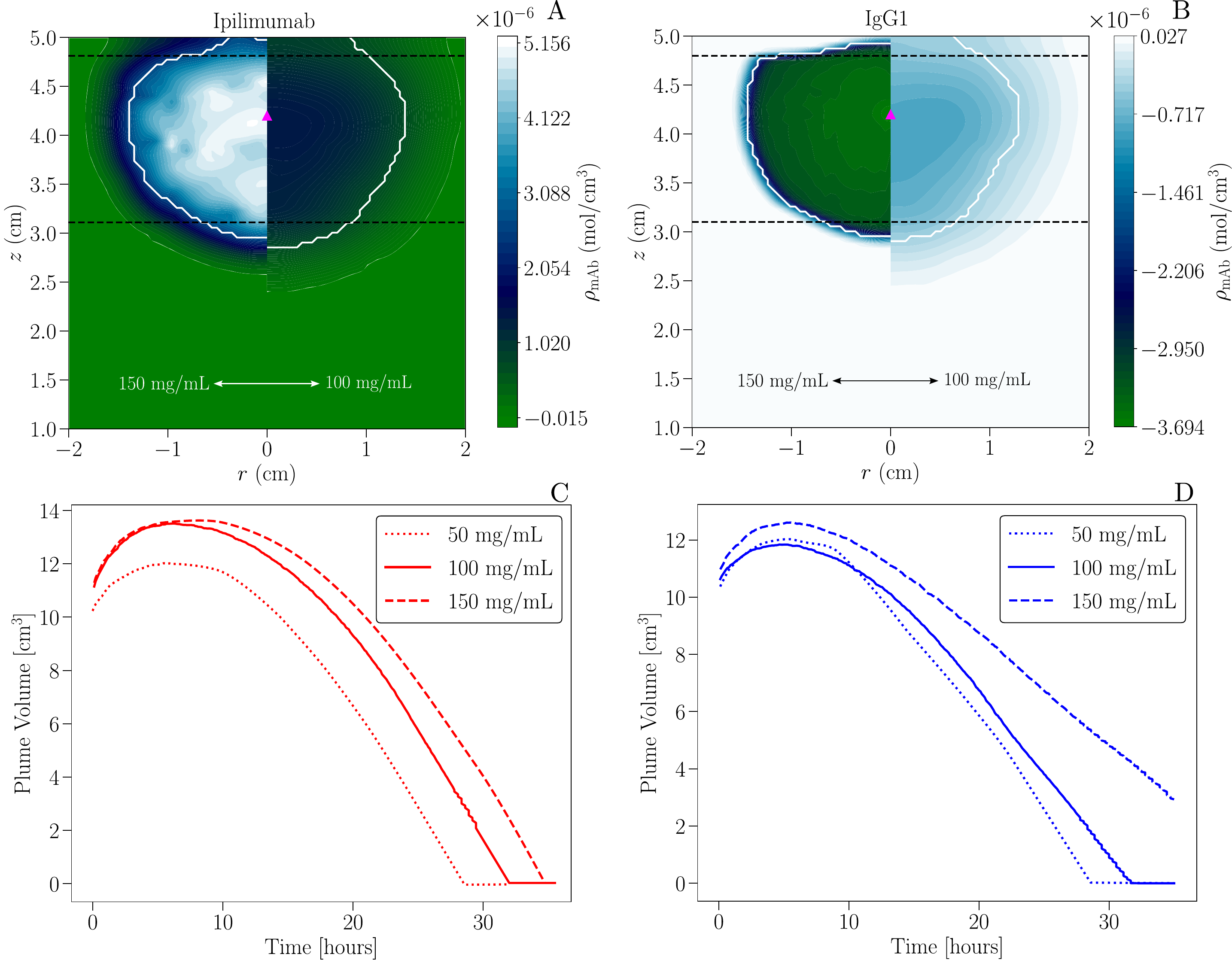}
    \caption{(A-B) Spatial distribution of net charge drug density at $t=20$ h for Ipilimumab and IgG1 across different concentrations. The white solid line, representing the isoline of $\text{sign} \left( c_{\text{mAb}} - 0.5 \max \left( c_{\text{mAb}} \right) \right)$, delineates the edge of the drug depot. The dashed black lines mark the tissue layer interfaces. The magenta triangle marks the injection point. (C-D) Plume volume over time for different concentrations.}
    \label{Charge_Density_conc}
\end{figure}

Figs.~\ref{Charge_Density_conc}A-B show the spatial distribution of net charge density at $t=20$ h for Ipilimumab and IgG1 at different formulation concentrations. For Ipilimumab, the net charge within the drug depot remains positive in both the low and high concentration cases. However, the maximum net charge in the high concentration formulation is approximately 3.38-fold higher than in the low concentration case. For IgG1, the net charge within the depot is negative under both conditions, with a maximum magnitude of $3.69 \times 10^{-6}$ for the high-concentration case compared to $8 \times 10^{-7}$ for the low-concentration case. Additionally, depots formed at higher concentrations remain more spread for both antibodies.

Figs.~\ref{Charge_Density_conc}C-D illustrate the time evolution of plume volume. In both cases, higher concentrations produce slightly larger plume volumes and extend the depot lifetime beyond 30 h, whereas lower concentrations show faster plume decay, with near-complete disappearance around 30 h. 
\subsubsection{Comparison with experiments}
For comparison with experiments, we use the experimental data from \cite{Stanton2001}, where lymphoscintigraphy was employed to measure the removal rate constant for $^{99m}$Tc-labelled human immunoglobulin G (hIgG) in the oedematous proximal forearm and the hand (finger web) in women. We selected the control data for the forearm and hand. 

\begin{figure}[h]
    \centering
    \includegraphics[width=0.99\columnwidth]{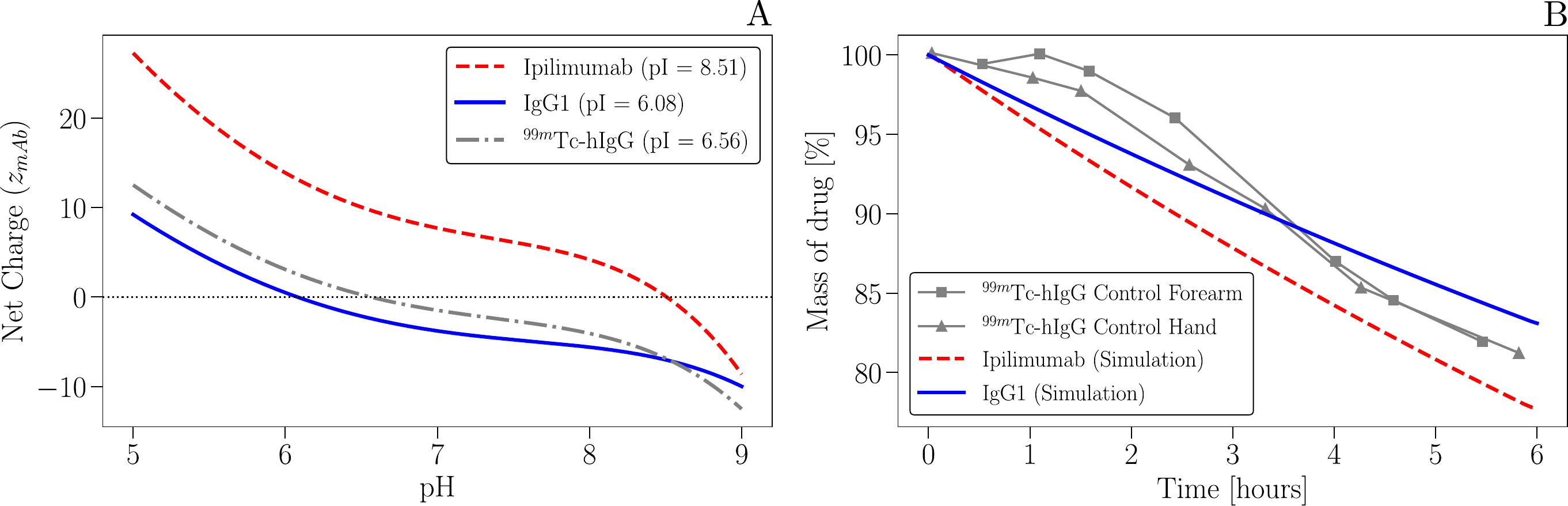}
    \caption{(A) Isoelectric plots of the mAbs used for comparison against experiments. 
    (B) Comparison of the averaged drug mass profiles across all buffer pH levels with depot clearance experiments. The experimental data correspond to $^{99m}$Tc-labelled human immunoglobulin G (hIgG) from \cite{Stanton2001}.}
    \label{experiments}
\end{figure}

Figure~\ref{experiments}A presents the isoelectric plots for all mAbs considered, showing that the mAb used in \cite{Stanton2001} exhibits a very similar isoelectric profile and isoelectric point to the IgG1 used in our simulations. The comparison with depot clearance experiments, shown in Figure~\ref{experiments}B, demonstrates good agreement between the experimental data and our simulations for both IgG1 and Ipilimumab across all pH levels considered. 

The simulated profiles capture the observed trends in drug clearance over time, with IgG1 exhibiting a slower decline in depot concentration compared to Ipilimumab, aligning more closely with the experimental values. This consistency suggests that the model accurately represents the dynamics of depot clearance for each mAb, capturing differences in clearance rates likely due to variations in binding affinity and tissue interactions.
\section{Conclusions}
Our findings offer valuable insights into the role of electric charge in mAb transport, binding, and absorption. In the short term, our simulations reveal that: (1) fluid flow generated by injection significantly impacts the initial transport of charged species, due to the predominating advective effects during the injection; (2) a higher buffer pH increases the electric potential after the end of the injection, resulting in increased tissue pH and reduced net charge drug density; (3) mAbs with a higher isoelectric point (pI) show higher charge density and lower average tissue pH. 

In the long term, mAbs with a higher pI exhibit higher lymphatic uptake and larger plume volumes. Buffer pH plays a critical role: acidic (pH $< 7.4$) pH levels increase lymphatic uptake for high-pI mAbs like Ipilimumab, while they also lead to larger plume volumes. At alkaline pH (pH $\geq 7.4$), pharmacokinetic differences between Ipilimumab and IgG1 are reduced, with both mAbs displaying similar absorption profiles and plume volumes.  

BMI strongly affects drug transport. A higher BMI increases lymphatic uptake for Ipilimumab, whereas IgG1 absorption remains similar across different BMI categories. A higher BMI also results in larger plume volumes that stay longer in the tissue. Regarding injection depth, we see that shallower injections increase lymphatic uptake for Ipilimumab.

Electromigration effects become more important at higher formulation concentrations, leading to more spread depots that last longer in the tissue, whereas lower concentrations disperse more rapidly, resulting in faster plume disappearance, and reduced lymphatic uptake.

Future studies should focus on enhancing the current computational model by incorporating tissue deformation. As demonstrated in \cite{DELUCIO2024124446}, tissue deformation significantly affects the shape and volume of the plume. Additionally, it is essential to generalize these findings across a wider range of mAbs, as this study considers only two types.





\bibliographystyle{unsrt}
\biboptions{sort&compress}
\bibliography{references}







\end{document}